\documentclass[english,floatfix,twocolumn,showkeys,superscriptaddress,nofootinbib,aps,prd]{revtex4}
\usepackage[latin1]{inputenc}
\usepackage[T1]{fontenc}
\usepackage{lmodern}
\usepackage{amsmath}
\usepackage{amssymb}
\usepackage{graphicx}
\usepackage{esint}
\usepackage{longtable}
\usepackage{dcolumn}
\usepackage{babel} 
\usepackage{csquotes} 
\usepackage{color} 
\usepackage{setspace}

\begin{document}

\title{Five-dimensional regular black holes in a brane world}

\author{Juliano C. S. Neves} 
\email{juliano.neves@unifal-mg.edu.br}
\affiliation{Instituto de Ciência e Tecnologia, Universidade Federal de Alfenas, \\ Rodovia José Aurélio Vilela,
11999, CEP 37715-400 Poços de Caldas, MG, Brazil}

\begin{abstract}
Following a recent approach, complete and analytic solutions  (brane and bulk) of regular black holes are shown
in a brane context. The metrics are regular both on the four-dimensional
 brane and in the five-dimensional bulk. Like many braneworld scenarios, the bulk spacetime is asymptotically 
 anti-de Sitter. On the other hand, a de Sitter core on the brane avoids the singularity inside the event horizon, 
 providing then well-known regular black holes on the brane. From the bulk perspective, the regular black holes are
five-dimensional objects, with the event horizon extending to the extra dimension,  
but the de Sitter core is entirely on the four-dimensional brane.
\end{abstract}


\maketitle

\section{Introduction}

The Randall-Sundrum model I  \cite{RSI} was conceived of as an attempt to solve the hierarchy problem in particle
physics by using a finite extra dimension and two branes (in which our universe is a brane) 
embedded into a five-dimensional spacetime. On the other hand, the Randall-Sundrum II 
 \cite{RSII} adopts just one brane, with an infinite extra dimension and a warped factor in the bulk
or five-dimensional spacetime that, in the second model version, confines gravity close to our universe. 
Since then such models were studied in other contexts like gravitation and 
cosmology.\footnote{See Refs. \cite{Maartens,Clifton} for a review on brane world scenario.}
The existence of either a finite or an infinite extra dimension in a context in which our 
four-dimensional world, a brane, is embedded into a
five-dimensional bulk creates new phenomena and has gained attention of many researchers in the last decades.

In gravitation, after the so-called black string from Hawking \textit{et al.} \cite{Hawking}, a large amount of
articles exploring the Randall-Sundrum models and the gravitational phenomenon in that context discusses
black holes and wormholes. We can divide the area into two approaches: from the brane or 
from the bulk. In the first one, black hole and wormhole geometries are built from the four-dimensional brane, 
and equations in the five-dimensional bulk are ignored \cite{Casadio:2001jg,Bronnikov:2003gx,Molina:2010yu,Molina:2012ay,Neves:2012it,Lobo:2007qi,Barcelo:2000ta,Dadhich:2000am,Aliev:2005bi,Neves:2015vga}. 
The second one builds spacetimes from the
bulk and tries to find out the brane equations \cite{Creek:2006je}. The approach adopted here is the second one. 

In a recent work, Nakas and Kanti \cite{Nakas} presented an approach in which the geometry in the bulk is
analytically obtained, and then the brane spacetime metric (a known solution in general relativity) and 
the energy-momentum tensor (an effective tensor, for example)
are fully calculated in a Randall-Sundrum II-type model. 
The main accomplishment of the Nakas-Kanti approach---from bulk to the brane---is
to obtain a well-known black hole solution on the brane, like the Schwarzschild geometry Ref. \cite{Nakas}.
This will be the approach adopted here. But instead of a black hole with a constant mass
like the Nakas and Kanti five-dimensional black holes \cite{Nakas,Nakas2},
 I use a mass function in order to generate a regular black hole (RBH) 
or a class of RBHs. Contrary to the black string, which is singular in the bulk, or the recent geometries of
Nakas and Kanti \cite{Nakas,Nakas2}, which are regular in the bulk but they are singular on the brane,
 the geometry (or geometries) presented here is regular in both spacetimes.
  In particular, the study of the model on the brane will be feasible due to the
induced gravitational field equations proposed by Shiromizu, Maeda, and Sasaki \cite{Shiromizu}. The induced field
equations provide an effective energy-momentum tensor on the brane. As we will see, the brane spacetime
is a vacuum spacetime, and the origin of the RBH on the brane is due to the bulk influence on the
brane. 

Like the Nakas and Kanti geometry \cite{Nakas}, the RBH presented here is a five-dimensional black hole, 
that is to say, its event 
horizon extends to the five-dimensional bulk. Moreover, in order to support a RBH, there is
a de Sitter core inside the event horizon, just on the brane spacetime, responsible for avoiding the
singularity issue. A de Sitter core is, for many RBH solutions, the main feature of such objects \cite{Bardeen,Ansoldi,Lemos:2011dq,Dymnikova:2003vt,Neves_Saa,Masa:2020hok}.

RBHs have been studied since the pioneer work of Bardeen \cite{Bardeen}.
 This class of black holes, like a good black hole,
presents horizons (an inner and an outer horizon) and precludes the central singularity. As I said, a de Sitter core
avoids the singularity, then the spacetime of RBHs is entirely regular. The reason for that is some violation of
energy conditions inside such objects. With energy conditions banned, the singularity theorems are not 
valid anymore.\footnote{See, for example, Ref. \cite{Wald}, chapter 9, for a detailed study on the singularity theorems.}
Thus the existence of a singularity inside a RBH is not a necessary consequence from a mathematical theorem.

The bibliography on RBHs has increased in the last years. Whether in the general relativity context \cite{Bardeen,Ansoldi,Lemos:2011dq,Dymnikova:2003vt,Neves_Saa,Masa:2020hok,Dymnikova:1992ux,Dymnikova:1996,Hayward,Neves:2015zia,Smailagic:2010nv,Bambi:2013ufa,Toshmatov:2014nya,Azreg-Ainou:2014pra,Neves_Saa2,Maluf_Neves2} or in
other contexts \cite{Bronnikov:2000vy,Modesto:2010rv,Ghosh:2018bxg}, RBHs have been studied. 
Even in the brane world context, there are articles that calculate and
explore spacetime metrics of RBHs \cite{Molina:2010yu,Neves:2015vga}. 
Here, the spacetime geometry is obtained both in the bulk and on
the brane. Contrary to previous articles, the complete and analytic solution is then presented, corresponding to 
 known four-dimensional geometries on the brane, like the Bardeen RBH.

This article is structured as follows: in Sec. II  one presents the Nakas and Kanti approach and applies it to RBHs,
generating then a five-dimensional RBH. In Sec. III the bulk energy-momentum tensor
is studied and is shown that energy conditions are violated. In Sec. IV the gravitational field equations are
fully calculated, and it is shown that, even with a four-dimensional RBH on the brane, this
spacetime is a vacuum spacetime. The final remarks are given in Sec. V. 

In this work, geometrized units are adopted. Then $G=c=1$ throughout this article. Capital Latin index runs 
from 0 to 4, and Greek index runs from 0 to 3.
 
\section{The bulk perspective}

\subsection{Five-dimensional regular black hole}
The five-dimensional metric in the Randall-Sundrum model II \cite{RSII} is given by
\begin{equation}
ds^2=e^{-2k\vert y \vert}\left(-dt^2 + d\vec{x}^2  \right) + dy^2,
\label{Metric1}
\end{equation}
where $k$ is related to the anti-de Sitter curvature radius $\ell_{AdS}$ by $k=1/\ell_{AdS}$. The five-dimensional
bulk $\mathcal{M}$ is asymptotically anti-de Sitter, and the four-dimensional brane $\Sigma$ is located at $y=0$.
In the flat coordinates, i.e., when the metric (\ref{Metric1}) is conformally flat, one has
\begin{equation}
ds^2=\frac{1}{\left(1+k\vert z \vert \right)^2}\left(-dt^2+dr^2+r^2d\Omega_2^2+dz^2 \right),
\label{Metric2}
\end{equation}
where the new coordinate $z=sgn (y)(e^{k\vert y\vert}-1)/k$ is adopted (note that the brane is still located at $z=0$), and $d\Omega_2^2=d\theta^2+\sin^2 \theta d\phi^2$ is the line-element of a unit two-sphere.

\begin{figure}
\begin{centering}
\includegraphics[scale=0.7]{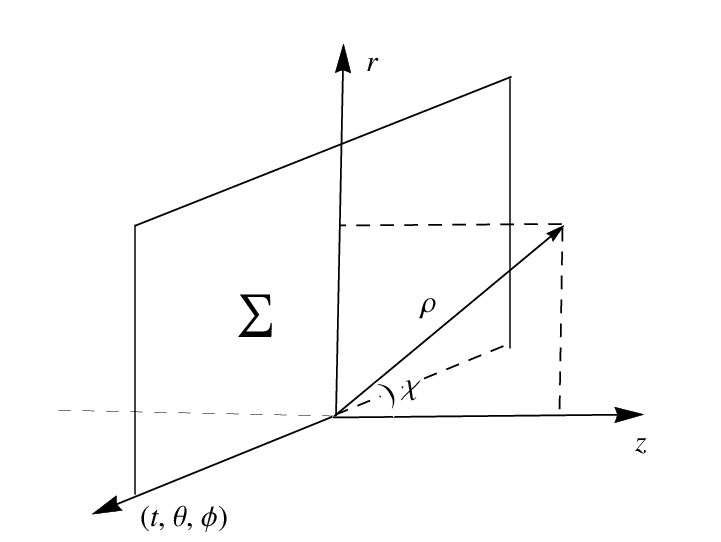}
\par\end{centering}
\caption{Brane world context: the four-dimensional brane is indicated as $\Sigma$, described by the
$(t,r,\theta,\phi)$ coordinates and is located at $z=0$. It is embedded into the five-dimensional bulk.}
\label{Diagram}
\end{figure}

As I said, the Nakas-Kanti  approach \cite{Nakas} to obtain both the bulk metric and the brane metric 
will be used here. According to
those authors, the next step from the general metric (\ref{Metric2}) regards to 
impose the spherical symmetry in the bulk. 
For that purpose, the following change of coordinates is necessary:
\begin{equation}
r= \rho \sin \chi  \hspace{0.5cm} \mbox{and} \hspace{0.5cm} z= \rho \cos \chi ,
\label{r,z}
\end{equation}
with $\chi \in [0,\pi]$. With those coordinate transformations, the metric (\ref{Metric2}) now is written as
\begin{equation}
ds^2=\frac{1}{\left(1+k\rho \vert \cos\chi \vert \right)^2}\left(-dt^2+d\rho^2+\rho^2d \Omega_3^2 \right),
\label{Metric3}
\end{equation}
where $d\Omega_3^2$ plays the role of the line-element from a unit three-sphere, that is to say,
\begin{equation}
d\Omega_3^2= d\chi^2+\sin^2 \chi d\theta^2+\sin^2 \chi \sin^2 \theta d\phi^2.
\end{equation}
And the inverse transformations of (\ref{r,z}) are given by
\begin{equation}
\rho =\left(r^2+z^2\right)^{\frac{1}{2}} \hspace{0.5cm} \mbox{and} \hspace{0.5cm} \tan \chi =\frac{r}{z}.
\label{rho}
\end{equation}
The new radial coordinate $\rho$ is a mix of the bulk extra coordinate and the brane radial coordinate $r$, it ranges
from 0 to $\infty$. The coordinate $\chi$ indicates the \enquote{left} and \enquote{right} side of the brane (see Fig. \ref{Diagram}). 
For $[0,\pi/2[$, one has the \enquote{right} side (positive values of $z$), and for $]\pi/2,\pi]$, the \enquote{left} one
(or negative values of $z$). Due to the $\mathbb{Z}_2$ symmetry in the bulk,  points $z$ and $-z$ are equivalent.  
Therefore, for some calculations, I will rule out the modulus of $y$ in Eq. (\ref{Metric3}).
 
Following the famous work on the black string \cite{Hawking}, Nakas and Kanti \cite{Nakas} replaced the part of the 
line element (\ref{Metric3}),
namely $-dt^2 + d\rho^2$, by the metric elements of the Schwarzschild geometry. Thus, the metric (\ref{Metric3})
 assumed the following form: 
\begin{equation}
ds^2=\frac{1}{\left(1+k\rho  \cos\chi  \right)^2}\left(-f(\rho) dt^2+\frac{d\rho^2}{f(\rho)} +\rho^2d \Omega_3^2 \right).
\label{Metric4}
\end{equation}
But in order to build a five-dimensional RBH (or an entire class of RBHs), I consider a mass function 
instead of a constant mass.
That is, the metric element $f(\rho)$ is then written as  
\begin{equation}
f(\rho)=1-\frac{2M(\rho)}{\rho}.
\label{f_rho}
\end{equation}
An appropriate mass function that produces RBHs (as we will see) can be written as
\begin{equation}
M(\rho)=\frac{M_0}{\left[1+\left(\frac{r_0}{\rho}\right)^q \right]^{\frac{3}{q}}}.
\label{Mass_function1}
\end{equation}
Such a mass function generated RBHs in the general relativity context. It is worth mentioning that
the mass function (\ref{Mass_function1}) provided solutions of Einstein's field equations by using
the so-called Synge g-method, in which from a given metric (with, for example, the mass
function (\ref{Mass_function1})), one obtains and interprets the 
resulting energy-momentum tensor (see Ref. \cite{Hernandez:1967zza} for more details on the Synge method).
Introduced in
Ref. \cite{Neves_Saa}, the mass function (\ref{Mass_function1}) is able to produce RBHs with and without rotation.
 In particular, for $q=2$ we
 have the Bardeen RBH \cite{Bardeen}, and for $q=3$ the Hayward RBH \cite{Hayward} is obtained. 
 The constant $M_0$ is
 conceived of as the black hole mass, and $q$ is a positive integer in the mass function. 
 On the other hand, $r_0$ is adequate to---at least---two interpretations.
For $q=2$, the constant $r_0$ is interpreted as a charge in a nonlinear electrodynamics according
 to Ayón-Beato and Garcia's work \cite{Beato}. 
 But in our work \cite{Maluf_Neves}, $r_0$ is conceived of as length, related to the Planck length.
 Even an upper bound on $r_0$ was assumed but in the
 general relativity context.\footnote{The upper bound obtained in Ref. \cite{Maluf_Neves} was $r_0<10^{-25}$m.} 
 However, as we will see, a mass function like (\ref{Mass_function1}) is also able to provide RBHs in a brane context
 assuming that $r_0$ is a short length. 

The capability of $M(\rho)$ removing the central singularity is made clear from the Ricci scalar $R$ and from
$R_{MN}R^{MN}$ and $R_{MNLK}R^{MNLK}$ (where the last one is also called Kretschmann scalar):
\begin{align}
& \lim_{\rho \rightarrow 0} R = -20k^2 +\frac{40M_0}{r_0^3}, \\
& \lim_{\rho \rightarrow 0} R_{MN}R^{MN} =  80k^4 -\frac{320M_0}{r_0^3}\left(k^2-\frac{M_0}{r_0^3} \right), \\
& \lim_{\rho \rightarrow 0} R_{MNLK}R^{MNLK} = 40k^4  -\frac{160M_0}{r_0^3}\left(k^2-\frac{M_0}{r_0^3} \right).
\end{align}    
As we can see, all results are $q$-independent. Also the metric elements are regular from that mass
function, i.e., $g_{tt}=-g_{rr}=-1$ and $g_{\theta\theta}=g_{\phi\phi}=g_{\chi \chi}=0$ for $\rho \rightarrow 0$. 

On the other hand, the anti-de Sitter behavior of (\ref{Metric4}) is indicated as $\rho \rightarrow \infty$. That is, 
\begin{align}
& \lim_{\rho \rightarrow \infty} R = -20k^2, \\
& \lim_{\rho \rightarrow \infty} R_{MN}R^{MN} =  80k^4, \\
& \lim_{\rho \rightarrow \infty} R_{MNLK}R^{MNLK} = 40k^4.
\end{align} 
Therefore, even with a mass function in Eq. (\ref{Metric4}), the bulk is still asymptotically anti-de Sitter.

\begin{figure}
\begin{centering}
\includegraphics[scale=0.53]{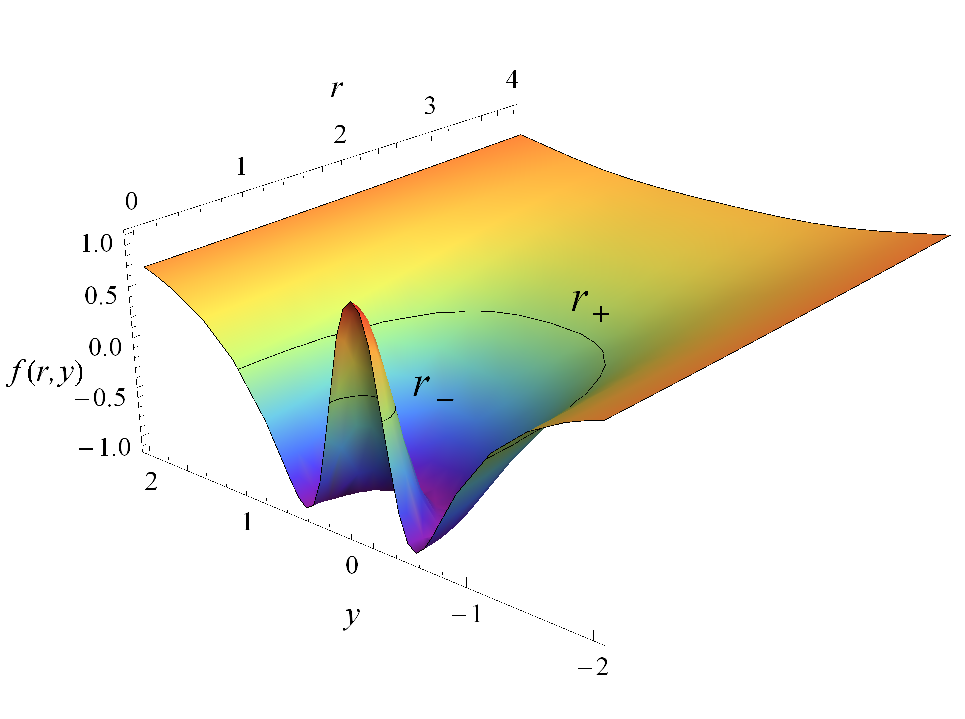}
\par\end{centering}
\caption{Zeros of $g^{rr}=f(r_{\pm},y_0)=0$ (indicated by the black lines) give us the localization of  horizons. 
The mass function (\ref{Mass_function2}) is able 
to provide two horizons, the inner $r_-$ and the outer horizon $r_+$. As we can see, the horizons get close to the 
brane (located at $y=0$). In this graphic, one uses $k=1$, $M_0=1,r_0=0.5$ and $q=3$.}
\label{Horizons}
\end{figure} 
 
In order to better illustrate the event horizon and then present the five-dimensional bulk RBH, one writes 
the metric (\ref{Metric4}) in the original $(t,r,\theta,\phi,y)$ coordinates. Thus
\begin{align}
 ds^2 &= e^{-2k\vert y \vert}\bigg[ -f(r,y)dt^2+\bigg(\frac{r^2}{f(r,y)}+z(y)^2 \bigg) \nonumber \\
& \times \frac{dr^2}{r^2+z(y)^2}+  \bigg(\frac{1}{f(r,y)}-1 \bigg)\frac{2rz(y)e^{k\vert y \vert}}{r^2+z(y)^2}drdy \nonumber \\ 
 & +  r^2d\Omega_2^2 \bigg] + \bigg(r^2+\frac{z(y)^2}{f(r,y)}\bigg)  \frac{dy^2}{r^2+z(y)^2},
\label{Metric5}
\end{align}
with
\begin{align}
f(r,y) & =1-\frac{2M(r,y)}{\left[r^2+z(y)^2\right]^{\frac{1}{2}}},\\
 M(r,y) & = M_0\left\{ \frac{\left[r^2+z(y)^2\right]^{\frac{3}{2}}}{\left(r_0^q+\left[r^2+z(y)^2\right]^{\frac{q}{2}} \right)^\frac{3}{q}}\right\},
 \label{Mass_function2}
\end{align}
and, as we saw, $z(y)=sgn(y)(e^{k\vert y \vert}-1)/k$. As pointed out by Nakas and Kanti \cite{Nakas}, contrary to the
 Hawking \textit{et al.} \cite{Hawking} black string, the metric (\ref{Metric5}) is nonfactorized. 
 And contrary to the geometry studied in Refs. \cite{Nakas,Nakas2}, there are two horizons here. 
 The mass function (\ref{Mass_function2})
 is able to provide at most two horizons, an inner $r_-$ and an outer horizon $r_+$, the event horizon properly 
 speaking, since $r_0<M_0$. Zeros of $g^{rr}(r_{\pm},y_0)=f(r_{\pm},y_0)=0$ give us the 
 localization of the horizons. As we can see  from Fig. \ref{Horizons}, both horizons extend to the bulk. 
 This very feature was pointed out by Nakas and Kanti \cite{Nakas,Nakas2} 
 and then was argued the \enquote{pancake} shape (for a fixed $y=y_0$) of the event horizon studied by the authors.  
 For a constant mass or $r_0=0$, the \enquote{pancake} shape of the horizon (in this case only one) is indicated by
 $r_+^2=4M_0^2-(e^{k\vert y_0 \vert}-1)^2/k^2$, and $y_0$ is simply obtained by making $r_+=0$, thus
 $y_0=\ln (2M_0k+1)/k$ for a constant mass. The value $y_0$ means the value of the extra coordinate in which
 $r_+$ shrinks exponentially and becomes zero. For our mass function, $y_0$ is obtained 
 from $2kM(0,y_0)-e^{k\vert y_0 \vert}+1=0$, which give us two values for the edge of the 
 horizons in the extra dimension, namely $y_{0 \pm}$ (related to the outer and inner horizons, respectively).

\section{The matter content in the bulk}
As we can see from the form of the metric element $f(\rho)$ adopted here, the five-dimensional bulk is
not an empty or flat spacetime. In order to illustrate this point, following Nakas and Kanti \cite{Nakas} and
specifying the bulk energy-momentum tensor, one writes the 
bulk gravitational action as
\begin{equation}
S_B=\int d^5x \sqrt{-g}\left(\frac{{}^{(5)}R}{2\kappa_5^2}+\mathcal{L}_m^{(B)} \right),
\label{Action}
\end{equation} 
in which $g$ is the metric determinant (from the bulk metric $g_{MN}$), ${}^{(5)}R$  is the five-dimensional
Ricci scalar, $\kappa_5^2=8\pi G_5$ is defined by the gravitational constant in five dimensions, and
$\mathcal{L}_m^{(B)}$ is the Lagrangian of the bulk matter content. By means of the variation of (\ref{Action}) 
with respect to $g_{MN}$, 
we have the field equations of the bulk, i.e.,
\begin{equation}
G_{MN}=\kappa_5^2 T_{MN}^{(B)}=-\frac{2\kappa_5^2}{\sqrt{-g}}\frac{\delta \left(\mathcal{L}_m^{(B)} \sqrt{-g} \right)}{\delta g^{MN}},
\end{equation}
where $G_{MN}$ is the five-dimensional Einstein tensor, and $T_{MN}^{(B)}$ is the bulk energy-momentum tensor. 
From the metric (\ref{Metric4}), the nonvanishing energy-momentum tensor components 
in the $(t,\rho,\theta,\phi,\chi)$ coordinates are given by
\begin{align}
 T^{(B)t}_{\ \ \ \ \  t} & = T^{(B)\rho}_{\ \ \ \ \  \rho} = \frac{1}{\kappa_5^2}\bigg[6k^2 + \bigg(\frac{9k \cos\chi }{\rho^2} - \frac{3}{\rho^3}\bigg) M(\rho)  \nonumber \\
& - \bigg( \frac{3k \cos \chi}{\rho}+\frac{3}{\rho^2} \bigg) M'(\rho) \bigg],
 \label{EM_tensor1}
\end{align}
\begin{align}
 T^{(B)\theta}_{\ \ \ \ \  \theta} & = T^{(B)\phi}_{\ \ \ \ \ \phi}= T^{(B)\chi}_{\ \ \ \ \  \chi}= \frac{1}{\kappa_5^2}\bigg[ 6k^2 - \bigg(\frac{6k^2 \cos^2 \chi }{\rho} \nonumber \\
 & -\frac{6k \cos \chi }{\rho^2}  \bigg) M(\rho) + \bigg(4k^2\cos^2 \chi +\frac{2k \cos \chi}{\rho} \nonumber \\
 & -\frac{2}{\rho^2}\bigg) M'(\rho) - \bigg(k^2 \rho \cos^2 \chi +2k \cos \chi  \nonumber \\
 &+\frac{1}{\rho} \bigg) M''(\rho) \bigg],  
 \label{EM_tensor2}
\end{align}
where the operator $'$ means derivative with respect to the function argument. 
As we can immediately read from Eqs. (\ref{EM_tensor1})-(\ref{EM_tensor2}), 
the bulk energy-momentum tensor is diagonal in 
the $(t,\rho,\theta,\phi,\chi)$ coordinates, it can be written
 as $T^{(B)\mu}_{\ \ \ \ \ \nu}=diag(-\rho_E, p_1, p_2,p_2,p_2)$, in which $\rho_E$ 
 is the energy density, and $p_1$ and $p_2$ are pressures. As $p_1\neq p_2$, thus the bulk spacetime
 is supported by an anisotropic fluid. Most importantly is the limit of the components of that tensor as $\rho \rightarrow \infty$. That is to say,
\begin{align}
\lim_{\rho \rightarrow \infty} \rho_E= & -\frac{6k^2}{\kappa_5^2}=\Lambda_{5} \\
\lim_{\rho \rightarrow \infty} p_1= & \lim_{\rho \rightarrow \infty} p_2= \frac{6k^2}{\kappa_5^2}= - \Lambda_{5},
\end{align}
 in which $\Lambda_5$ is the five dimensional cosmological constant. 
 Therefore, the anti-de Sitter feature of the bulk gets evident once again. 
 
\begin{figure}
\begin{centering}
\includegraphics[trim=0cm 0cm 2cm 0cm, clip=true,scale=0.55]{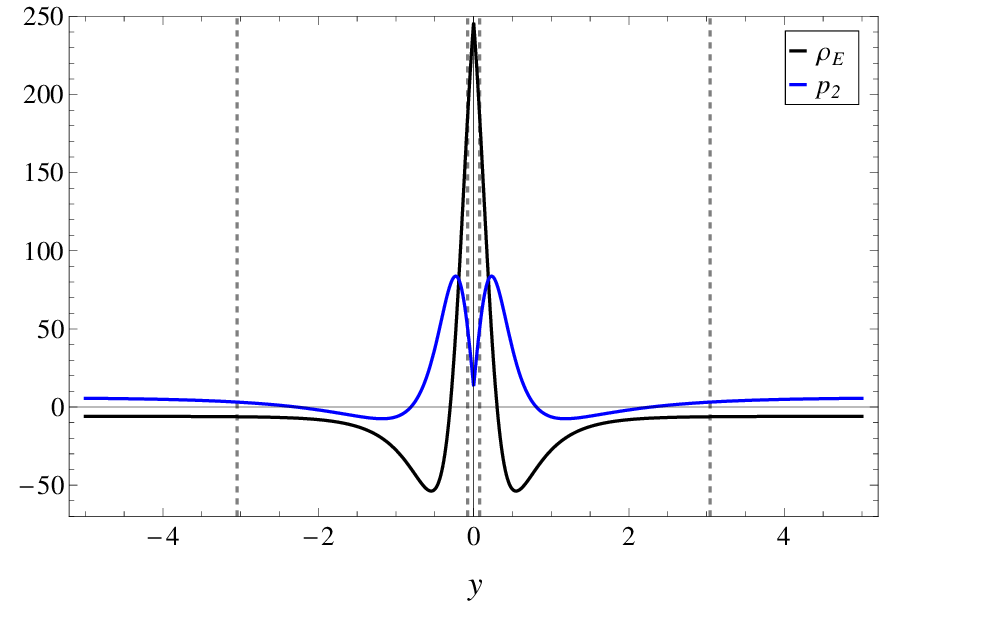}
\par\end{centering}
\caption{Components of the bulk energy-momentum tensor and the weak energy violation close to the brane, located
 at $y=0$. Vertical dashed lines indicate the horizons edges in the bulk (inner $y_{0-}=\pm 0.08$ and 
 outer $y_{0+}=\pm 3.04$). In this graphic, one uses $\kappa_5=k=1$, $M_0=10,r_0=0.1$, $q=2$ and, 
 most importantly, $r=0.5$.}
\label{Energy_conditions1}
\end{figure}

\begin{figure}
\begin{centering}
\includegraphics[trim=0.3cm 0cm 0.6cm 0cm, clip=true,scale=0.57]{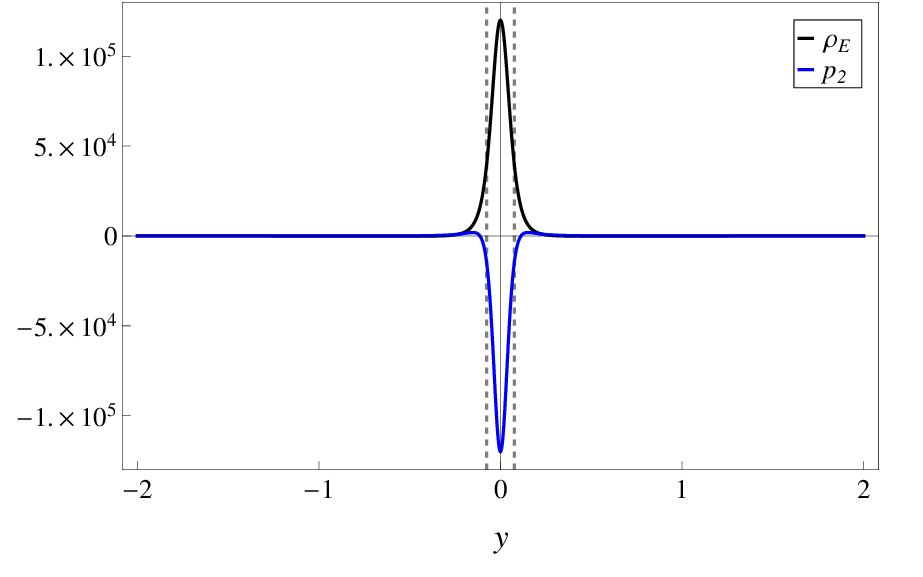}
\par\end{centering}
\caption{Components of the bulk energy-momentum tensor for $r=0$ on the brane. 
The vertical dashed lines indicate the inner horizon limit in the bulk ($y_{0-}=\pm 0.08$).  In this graphic, 
one uses $\kappa_5=k=1$, $M_0=10,r_0=0.1$ and $q=2$. For $y=0$, we see a de Sitter equation of state ($p_2/\rho_E=-1$).}
\label{Energy_conditions2}
\end{figure}

In order to see the dependence of (\ref{EM_tensor1})-(\ref{EM_tensor2}) on the extra coordinate, let us write $T^{(B)\mu}_{\ \ \ \ \ \nu}$ in the
$(t,r,\theta,\phi,y)$ coordinates. Using the transformations (\ref{rho}), we have   
\begin{align}
\rho_E& =-\frac{3k^2}{\kappa_5^2}\Bigg\{ 2 - \left(1+\delta \right)^{-\frac{3+q}{q}} \frac{ k M_0\left[ 4\left(1+\delta \right) -3e^{k\vert y \vert}\right]}{\left[k^2 r^2 + \left(e^{k\vert y \vert}-1 \right)^2 \right]^{\frac{3}{2}}} \Bigg\}, \\
p_2& = \frac{6k^2}{\kappa_5^2} \Bigg \{1+\frac{kM_0 \left(e^{k\vert y \vert}-1 \right) \left(2-e^{k\vert y \vert} \right)}{\left[k^2 r^2 + \left(e^{k\vert y \vert}-1 \right)^2 \right]^{\frac{3}{2}}} \nonumber \\
& - \frac{kM_0\left[\left(4-3e^{k\vert y \vert} - \left(1+q \right) \frac{1}{2}e^{2k\vert y \vert}  \right) \delta +4\delta^2 \right]}{\left[k^2 r^2 + \left(e^{k\vert y \vert}-1 \right)^2 \right]^{\frac{3}{2}}} \Bigg \} ,
\end{align}
with 
\begin{equation}
\delta = \left(\frac{k r_0}{\left[k^2r^2+\left(e^{k\vert y\vert}-1 \right)^2\right]^{\frac{1}{2}}} \right)^q.
\end{equation}
As we can see, for $\delta=0$ or, equivalently, $r_0 = 0$ we have the same results of Ref. \cite{Nakas}, 
and the mass function (\ref{Mass_function2}) turns into a constant. In the  $(t,r,\theta,\phi,y)$ coordinates, the
energy-momentum tensor is not diagonal anymore. Two new off-diagonal components $T^{(B)r}_{\ \ \ \ \  y}= e^{2k\vert y \vert}T^{(B)y}_{\ \ \ \ \  r}$ 
appear, but both are zero as $y \rightarrow \infty$, as we would expect for an asymptotically anti-de Sitter spacetime.
Other components will not be shown here due to the large and cumbersome final forms. The main idea here is just
indicate some energy condition violation for the five-dimensional bulk. In particular, as we can see in 
Fig. \ref{Energy_conditions1}, the weak energy condition is violated close to the brane, i.e., 
$\rho_E <0$ and $\rho_E + p_2 <0$.
Moreover, Fig. \ref{Energy_conditions2} shows that for $r=0$ and $y=0$, $p_2/\rho_E =-1$, 
which is the equation of state of a de Sitter spacetime. As we will see, this indicates that a de Sitter 
core is inside the event horizon and it is entirely on the brane, not in the bulk. For a five-dimensional de Sitter 
space one has $g_{tt}=-(1-r^2/\alpha^2)$, with $\alpha$ constant, and from Eq. (\ref{Metric5}), for small values of  the
radial coordinate $r$, we have $g_{tt}\simeq-(1-\beta r^2)$, with $\beta$ constant, only for $y=0$, that is, 
outside the extra dimension.

\section{The brane perspective}

\subsection{Gravitational field equations on the brane}
With $y=z=0$ in the metric (\ref{Metric5}), we have a four-dimensional RBH. But would it be a RBH in a brane context?
For sure in that context the gravitational field equations are not the Einsteinian equations. 
In the four-dimensional world, the brane, the field equations are very different ones. They were deduced from
the  Gauss and Codacci equations, according to the seminal work of Shiromizu, Maeda, and Sasaki \cite{Shiromizu}. 
I adopt here a scenario like that one. In such a scenario, the four-dimensional brane $(\Sigma, h_{MN})$ is
embedded into a five-dimensional bulk $(\mathcal{M},g_{MN})$, and a normal and unit vector to the brane
is written as $n^M= h^M_{\ y}$. Therefore, the induced metric on the brane reads $h_{MN}=g_{MN}-n_{M}n_{N}$.

Following Nakas and Kanti \cite{Nakas,Nakas2}, one defines the total energy-momentum tensor as
\begin{equation}
T_{MN}=T^{(B)}_{MN}+h^{\mu}_{\ M} h^{\nu}_{\ N}T^{(br)}_{\mu\nu} \delta (y),
\label{T_total}
\end{equation}
with the brane energy-momentum tensor, like many brane world scenarios, given by
\begin{equation}
T^{(br)}=-\lambda h_{\mu\nu}+\tau_{\mu\nu}.
\label{T_brane}
\end{equation}
The constant $\lambda$ is the brane tension, some sort of vacuum energy on the brane, and $\tau_{\mu\nu}$ 
regards all matter fields on the four-dimensional spacetime. As we will see, 
the tensor $\tau_{\mu\nu}$ vanishes, and the RBH on the brane will be supported by the bulk geometrical 
influence on the brane. Another important point here is that the total energy-momentum tensor (\ref{T_total}) is
different from that one chosen in Ref. \cite{Shiromizu}. The bulk energy-momentum tensor is not described by a 
cosmological constant term. As we saw in the previous section, an exotic field ensures the asymptotic 
anti-de Sitter behavior for the bulk spacetime instead of a five dimensional cosmological constant.

In order to show that $\tau_{\mu\nu}=0$ in the brane context studied here, 
one uses Israel's junction condition \cite{Israel:1966rt} at $y=0$ and a very important result 
for the extrinsic curvature $K_{\mu\nu}$, namely  
\begin{equation}
\left[K_{\mu\nu} \right]=-\kappa_5^2 \left(T^{(br)}_{\mu\nu} -\frac{1}{3}h_{\mu\nu}T^{(br)} \right),
\label{K1}
\end{equation}
where $T^{(br)}$ is the trace of the energy-momentum
tensor on the brane. The extrinsic curvature is defined as
\begin{equation}
K_{\mu\nu}= h^A_{\ \mu} h^B_{\ \nu} \nabla_A n_B,
\label{K2}
\end{equation}
with $n^{A}=(0,0,0,0,1/ \sqrt{g_{yy}(y=0)})$, and the bracket notation means
\begin{equation}
\left[X \right]= \lim_{y\rightarrow 0^+}X- \lim_{y \rightarrow 0^-}X=X^+-X^-.
\end{equation}
The induced metric $h_{\mu\nu}$ on the brane in the $(t,r,\theta,\phi)$ coordinates is written as
\begin{equation}
ds^2=-\left(1-\frac{2m(r)}{r} \right) dt^2+\left(1-\frac{2m(r)}{r} \right) ^{-1}dr^2+r^2d\Omega_2^2,
\label{Induced_metric}
\end{equation}
with
\begin{equation}
m(r)=M(r,0),
\label{Mass-brane}
\end{equation}
 and it clearly satisfies $[h_{\mu\nu}]=0$. With the aid of (\ref{Metric5}), 
then the extrinsic curvature reads simply 
\begin{equation}
K_{\mu\nu}=-k\frac{d\vert y \vert}{dy}h_{\mu\nu},
\label{K}
\end{equation}  
and follows that $K=-4k \frac{d\vert y \vert}{dy}$. With the extrinsic curvature calculated, the definition (\ref{K1})
(contracted with $h^{\mu\nu}$) leads to
\begin{equation}
T^{(br)}=\frac{3}{\kappa_5^2}[K].
\end{equation}
Therefore, Eq (\ref{K1}) can be rewritten as
\begin{equation}
T^{(br)}_{\mu\nu}=-\frac{1}{\kappa_5^2}\left(\left[K_{\mu\nu} \right]-\left[ K\right]h_{\mu\nu} \right)=-\frac{6k}{\kappa_5^2}h_{\mu\nu},
\end{equation} 
in which the $\mathbb{Z}_2$ symmetry  was used. From the above result and Eq. (\ref{T_brane}),
one concludes that $\tau_{\mu\nu}=0$, that is to say, \textit{there are no sources or matter fields on the brane}. Besides, 
the brane tension is related to the anti-de Sitter curvature by means of $\lambda=6k/\kappa_5^2>0$.
Then the RBH geometry should be produced by the bulk influence on the brane (this point will be emphasized 
again below). 

In order to describe the bulk influence on the brane,  the field equations induced on the brane are necessary, equations
with components that deforms the brane generating then a black hole. As I said, the induced field equations
were deduced by Shiromizu, Maeda, and Sasaki \cite{Shiromizu}. Such field equations are given by
\begin{align}
G_{\mu\nu}=& \frac{2\kappa_5^2}{3}\left(h^{M}_{\ \mu} h^{N}_{\ \nu} T^{(B)}_{MN}+\left[ T^{(B)}_{MN}n^M n^N -\frac{T^{(B)}}{4}  \right] h_{\mu\nu} \right) \nonumber \\
&  + KK_{\mu\nu}- K^{\ \alpha}_{\mu}K_{\nu \alpha}-\frac{1}{2}h_{\mu\nu}\left(K^2-K^{\alpha \beta}K_{\alpha \beta} \right) \nonumber \\
& - E_{\mu\nu},
\label{Field_equations}
\end{align}
where $G_{\mu\nu}$ is the four-dimensional Einstein tensor, $T^{(B)}$ is the trace of the 
bulk energy-momentum tensor, and $E_{\mu\nu}$ is the so-called \enquote{electric} part
of the five-dimensional Weyl tensor $(C^A_{\ \ BCD})$ projected onto the brane. Its form is
\begin{equation}
E_{\mu\nu}= C^A_{\ \ BCD}n_An^C h^{B}_{\mu} h^{D}_{\nu}.
\label{E_definition}
\end{equation}
It is worth pointing out that $E^{\mu}_{\ \mu}=0$, i.e., it is a traceless tensor.  Therefore, from the metric (\ref{Metric5}), with 
the mass function (\ref{Mass_function2}), $E_{\mu\nu}$ reads
\begin{equation}
E_{\mu\nu} \bigg|_{y \rightarrow 0 }= \frac{M_0  \mathcal{E}}{r^3} \left(\begin{array}{cccc}
-h_{tt}\\
 &-h_{rr} \\
 &  & h_{\theta \theta} \\
 &  &  & h_{\phi \phi}
\end{array}\right),
\label{E}
\end{equation}
 with
 \begin{equation}
\mathcal{E}= \frac{\left[2-\left(1+q \right)\left(\frac{r_0}{r} \right)^q \right]}{2 \left[1+ \left(\frac{r_0}{r} \right)^q \right]^{\frac{2q+3}{q}}}. 
 \end{equation}
For $r_0=0$, $M(r,0)=m(r)=M_0$ and $\mathcal{E}=1$, then we recover the results of Ref. \cite{Nakas}. Moreover,
 $E_{\mu\nu}$ is also regular due to the mass function, that is to say, $\lim_{r \rightarrow 0} E_{\mu\nu}=0$.

The terms of the gravitational field equations (\ref{Field_equations}) that contain the extrinsic 
curvature, with the aid of Eq. (\ref{K}), result in
\begin{align}
& KK_{\mu\nu}- K^{\ \alpha}_{\mu}K_{\nu \alpha}-\frac{1}{2}h_{\mu\nu}\left(K^2-K^{\alpha \beta}K_{\alpha \beta} \right) = 8\pi G \tau_{\mu\nu}  \nonumber \\
&+\kappa_5^4 \left( \pi_{\mu\nu}-\frac{\lambda^2}{12} h_{\mu\nu} \right),
\label{KK}
\end{align} 
with $G=\kappa_5^4 \lambda/48\pi=1$ playing the role of the effective 
gravitational constant on the brane, and
\begin{equation}
\pi_{\mu\nu}=-\frac{1}{4}\tau_{\mu\alpha} \tau^{\ \alpha}_{\nu}+\frac{1}{12}\tau \tau_{\mu\nu}+\frac{1}{8}\tau_{\alpha \beta}\tau^{\alpha \beta} h_{\mu\nu} -\frac{\tau^2}{24} h_{\mu\nu}.
\label{Tau}
\end{equation}
By using the above results, namely Eqs. (\ref{KK})-(\ref{Tau}), we are able to rewrite the field equations 
in terms of an effective energy-momentum tensor, i.e., 
\begin{equation}
G_{\mu\nu}=8\pi G \left(T^{(eff)}_{\mu\nu}+\tau_{\mu\nu} \right)+\kappa_5^4 \left(\pi_{\mu\nu}-\frac{\lambda^2}{12}h_{\mu\nu} \right) - E_{\mu\nu}.
\label{Field_eq2}
\end{equation}
It is worth emphasizing that $\pi_{\mu\nu}=0$ because, as we saw, $\tau_{\mu\nu}=0.$ The above field equations then
present an effective energy-momentum tensor (diagonal tensor) given by
\begin{equation}
T^{(eff)}_{\mu\nu}= \frac{2}{3k}\left(T^{(B)}_{\mu\nu}+\left[ T^{(B)}_{yy} -\frac{T^{(B)}}{4}  \right] h_{\mu\nu} \right) \Bigg|_{y=0},
\label{T_eff}
\end{equation}
calculated at $y=0$. The explicit form of (\ref{T_eff}) is then
\begin{align}
T^{(eff)}_{\mu\nu} = & \frac{1}{\kappa_5^2 k}\Bigg[ \frac{M_0}{r^3} \left(\begin{array}{cccc}
-\mathcal{T}_1 h_{tt}\\
 &- \mathcal{T}_1 h_{rr} \\
 &  & \mathcal{T}_2 h_{\theta \theta} \\
 &  &  & \mathcal{T}_2 h_{\phi \phi}
\end{array}\right) \nonumber \\
& + 3k^2 h_{\mu\nu} \Bigg],
\label{T_eff2}
\end{align}
with
\begin{align}
\mathcal{T}_1 & =  \frac{\left[12\left(\frac{r_0}{r}\right)^{2q}+ \left(11-q \right)\left(\frac{r_0}{r}\right)^{q} +2 \right]}{2\left[1+\left(\frac{r_0}{r}\right)^{q} \right]^{\frac{2q+3}{q}}}, \\
\mathcal{T}_2 &  = - \frac{\left[12\left(\frac{r_0}{r}\right)^{2q}- 5\left(1+q \right)\left(\frac{r_0}{r}\right)^{q} -2 \right]}{2\left[1+\left(\frac{r_0}{r}\right)^{q} \right]^{\frac{2q+3}{q}}}.
\end{align}
Once again, for $r_0=0$, $\mathcal{T}_1=\mathcal{T}_2=1$, and we recover the Nakas and Kanti 
results \cite{Nakas} when the mass parameter is a positive constant.

\subsection{Four-dimensional regular black hole}
The effective energy-momentum tensor (\ref{T_eff2}) is also regular on the brane. Calculating it, one has
\begin{equation}
\lim_{r \rightarrow 0}T^{(eff) \mu}_{\ \ \ \ \  \ \ \nu}=\frac{3}{\kappa_5^2}\left(k-\frac{2M_0}{k r_0^3} \right),
\label{Limit_T}
\end{equation} 
for all nonzero components. Assuming the field equations (\ref{Field_eq2}) for $r\rightarrow 0$,
we have $G^{\mu}_{\ \nu}=-\frac{6M_0}{r_0^3} h^{\mu}_{\ \nu}$, suggesting then a de Sitter core, 
that which prevents the singularity issue, inside the hole. Thus
\begin{equation}
\ell_{dS}= \left(\frac{r_0^3}{2M_0}\right)^{\frac{1}{2}},
\end{equation}
with $\ell_{dS}$ playing the role of the de Sitter curvature radius.
In this case, we have a de Sitter core inside the event horizon 
like well-known RBHs (as pointed out in Ref. \cite{Neves_Saa}). 

As we know, the gravitational field equations can provide the four-dimensional Einstein tensor. 
Calculating the right side of (\ref{Field_eq2}),
one has
\begin{equation}
G_{\mu\nu} =  - \left(\begin{array}{cccc}
 \frac{2m'(r) h_{tt}}{r^2}\\
 & \frac{2m'(r)h_{rr}}{r^2}  \\
 &  &  \frac{m''(r)h_{\theta \theta}}{r}  \\
 &  &  &  \frac{m''(r)h_{\phi \phi}}{r} 
\end{array}\right),
\label{G}
\end{equation}
which is the same Einstein tensor for a RBH in the general relativity context with $m(r)$ given by Eq. (\ref{Mass-brane}). 
But as our brane is a vacuum spacetime, the origin or source for that RBHs comes from the five-dimensional
bulk and its geometrical influence on the brane. With the Einstein tensor indicated above, we have then
well-known RBHs on the brane, whether the Bardeen RBH (for $q=2$) or the Hayward RBH (for $q=3$).

\section{Final remarks}
When applied to the gravitational phenomenon, brane-world models are richer than the Einsteinian context.
Like in the general relativity context, brane RBHs are even possible. Here a recent approach, due to
Nakas and Kanti \cite{Nakas}, was adopted. From the bulk spacetime to the brane spacetime (calculating
the induced gravitational field equations), the geometries presented here are regular both in the five-dimensional and 
on the four-dimensional world, contrary to the so-called black string or even the Nakas and Kanti geometries, which
are singular on the brane. Above all, the geometries obtained in this article are five-dimensional RBHs 
because their horizons extend to the extra dimension,
and, interestingly, the de Sitter core---that which precludes the central singularity---is entirely on the four-dimensional
brane.

From the induced field equations on the brane, we saw that the four-dimensional brane is a vacuum spacetime.
Thus, the source of the RBH on the brane is in the five-dimensional bulk. Then the bulk influence on the
brane is geometrical one. An important point in this article regards well-known RBHs in the 
general relativity context, like the Bardeen and Hayward RBH, which are found in the brane context adopted here.


\end{document}